\documentstyle[psfig]{mn}

\newcommand{\kms}	{km~s$^{-1}$}
\newcommand{\h}   	{$h^{-1}\,$~kpc}
\newcommand{\mpc}       {$h^{-1}\,$~Mpc}
\newcommand{\etal} 	{{et~al.}}
\newcommand{\push}[1]	{\multicolumn{1}{c}{#1}}
\newcommand{\hst}	{{\it{HST}}}
\newcommand{\lya}       {Ly$\alpha$}

\title[Ly$\alpha$ absorption at low-redshift]{
Ly$\alpha$ absorption in the nearby universe: the sightline to Q1821+643}

\author[D.V. Bowen, M. Pettini]
{David V. Bowen$^{1}$,
Max Pettini$^2$, and Brian J. Boyle$^3$\\
$^1$ Royal Observatory, Edinburgh, Blackford Hill, Edinburgh EH9 3HJ\\
$^2$ Royal Greenwich Observatory, Madingley Rd., Cambridge CB3 0EZ\\
$^3$ Anglo-Australian Observatory, P.O. Box 296, Epping, NSW 2121, Australia 
}

\date{Recieved 23 July 1997; accepted 10 December 1997}

\begin{document}

\maketitle

\begin{abstract}

We present the first results of a survey designed to understand the origin 
of the \lya -forest absorption systems at low redshift.  Using the 
WYFFOS and HYDRA multi-fibre spectrographs on the WHT and WIYN 
telescopes, we have identified 51 galaxies brighter than $b_j\:=\:18.5$ 
within 30 arcmins of the sightline of the QSO 1821+643.  
We find three galaxies within 500~\h\ of the QSO sightline; the nearest
galaxy is 104~\h\ away from the line of sight, and is at the same redshift as
a strong ($W_r\:=\:0.63$~\AA ) \lya\ absorption line. The remaining two
galaxies have no corresponding absorption to extremely low equivalent width
limits ($<0.05$~\AA ). Beyond 500~\h\ \lya\ absorption lines are found at
similar redshifts as several galaxies, but we show that these coincidences are
likely to be accidental.

Half of the \lya\ systems for which we could have found at least an $L^*$
galaxy have no galaxies at the absorbers' redshifts.  For the majority of the
remainder, we show that any apparent association with galaxies is probably
coincidental.  These \lya\ systems are characterised by their weak equivalent
widths ($W_r\:<\:0.2$~\AA ), and we conclude that this population of absorbers
is uncorrelated, or at best, weakly correlated, with galaxies.

\end{abstract}


\section{INTRODUCTION}

Until recently, it was widely believed that the numerous, diffuse H~I clouds
which give rise to the dense Ly$\alpha$-forest lines seen in high-redshift QSO
spectra are {\it intergalactic} clouds not directly associated with galaxies
(Sargent~et al.\ 1980).  However, since the launch of \hst\ and the subsequent
discovery of Ly$\alpha$ lines in the UV at redshifts $<$ 1 (Bahcall~et al.\
1991; Morris~et al.\ 1991), our understanding of their origin has become
somewhat confused.
The low-redshifts of these newly identified systems have made
it possible to directly search for galaxies which may be responsible for the
absorption, and deep CCD images of fields around observed QSOs, combined with
spectroscopic confirmation of candidate galaxies, initially suggested that
Ly$\alpha$ clouds are only weakly associated with galaxies on scales of
$\sim\:0.5-1$~Mpc (Morris~et al.~1993, hereafter M93; Stocke et
al. 1995). Later studies, however, found that most luminous galaxies at $0.1 <
z < 1$ are apparently surrounded by extended gaseous envelopes $\sim 160 \:
h^{-1}$~kpc in radius\footnote{$h\:=\:H_0/100$, where 
$H_0$ is the Hubble constant,
and $q_0\:=\:0.5$ throughout this paper} 
and that the fraction of Ly$\alpha$-forest lines which
occur in galaxies is high, between $\sim 40-70$~\% (Lanzetta~et al. 1995,
hereafter LBTW).  Confusingly, a study identical in method and objective, but
using different sightlines, failed to confirm this result (Le
Brun~et~al. 1996) and instead suggested that \lya\ lines may simply follow the
large scale structures which the brightest galaxies map out. More
recent observations have shown that at least some \lya\ absorbers reside
in galaxy voids (Shull~\etal\ 1996), although what that fraction is remains
unclear.

In a previous paper (Bowen, Blades, \& Pettini 1996; hereafter BBP) we
conducted a search for \lya\ absorption lines from {\it nearby} galaxies,
since it is easier to study the origin and characteristics of \lya\ clouds in
the local universe. 
We concluded that nearby galaxies do
not possess \lya -absorbing halos beyond $\simeq\:300$~\h , and that {\it if}
absorption systems were directly associated with chosen galaxies, then
$\sim 40$\% of galaxies had absorbing halos of radii $50-300$~\h . However,
several facts led us to question whether galaxies we identified were
actually responsible for a \lya\ line when absorption was found, 
and we proposed
that the most likely explanation was that absorbers follow the large-scale
structure of the galaxy distribution, a conclusion similar to that drawn by Le
Brun~\etal\ (1996) at higher redshift and supported by numerical simulations
of the high redshift universe (e.g.  Cen~\etal\ 1994; Petitjean, M\"{u}cket \&
Kates 1995; Zhang~\etal\ 1995; M\"{u}cket~\etal\ 1996).  An alternative
explanation is that \lya\ absorbers are associated with dwarf galaxies or low
surface-brightness galaxies; in that case the bright galaxies we had selected
which appeared to show absorption were not in fact directly responsible for
the absorption. A similar suggestion has been made by Shull~\etal\ (1996).

The observations reported in BBP were 
designed to study the incidence of absorption in galactic
halos; we did not, however, quantify the fraction of the
\lya -forest associated with galaxies. To determine this quantity, complete
redshift information must be obtained in the field of a QSO 
sightline to ensure that any galaxy which might be associated with any 
particular absorption line could be detected --- at least to some limiting 
magnitude.

We have therefore begun a spectroscopic survey of galaxies in fields of
selected QSOs observed with \hst\ in the far UV, to characterise 
any association
of \lya\ absorption systems with present day galaxies and their large scale
structure. With the advent of multi-fibre spectrographs able to obtain
redshifts of relatively faint galaxies over large areas of sky, it is now
feasible to pursue such a study.  In this paper we present the first results
of our programme, identification of a magnitude limited sample of galaxies
toward the bright QSO 1821+643 ($z_{\rm{em}}\:=\:0.297$). This was one of the
first QSOs to be observed with \hst\ (Bahcall~\etal\ 1992), and was one of
the sightlines studied by Le~Brun~\etal\ (1996) in order to identify
intermediate-redshift \lya\ absorbers. 
Q1821+643 is unique in being one of the few QSOs bright enough to be
observed with the Faint Object Spectrograph (FOS) aboard \hst\ {\it and} the
medium resolution gratings of the 
Goddard High Resolution Spectrograph (GHRS). This means that over small
wavelength regions it is possible to search for weak \lya\ lines (equivalent
widths $<\:100$~m\AA ) which cannot be seen in FOS data.

This paper is structured as follows: In \S2.1 we present spectra of galaxies
obtained with the multi-fibre spectrographs WYFFOS/AF-2 attached to the
William Herschel Telescope (WHT) on La Palma, and HYDRA on the WIYN telescope
at the Kitt Peak National Observatory.  
\S2.2 discusses our re-analysis of the avilable \hst\ data.  In
\S3 we present the results from our survey, and discuss the size of
individual galaxy halos (\S3.1) and the association of \lya\ lines with
galaxies in general (\S3.2). In \S3.3 we consider whether the associations
found are significant compared to a random distribution of galaxies and
absorbers.
Our conclusions are summarised in \S3.5.

\section{OBSERVATIONS}

\subsection{Galaxy redshifts}

\subsubsection{WYFFOS data}

Lists of galaxy candidates were produced from a scan of the POSS-II plate
PJ3948c which covers the field of Q1821+643 using the Automated Plate
Measuring (APM) Scanning Machine in Cambridge.  The majority of spectra were
obtained with the WYFFOS/AF-2 instrument on the WHT on August 20 and August
22nd 1995.  WYFFOS is a bench spectrograph situated at the Nasmyth platform of
the WHT, fed by 2.7 arcsec fibres positioned at the prime focus of the
telescope using the Autofib-2 fibre positioner, with a circular field of view
of 30 arcmins radius.  Data were obtained using the R600B grating centered at
5200~\AA\ and a thinned Tektronix 1024 CCD as the detector giving a wavelength
coverage of $\approx 3000$~\AA .

Fibre~configurations were produced using the {\tt configure} routine, with
priority given to the brightest objects in our sample. With a configuration
completed, objects already allocated with a fibre were removed from the list,
and the configuration routine run again, with the same magnitude priority
acting on the list to produce the next configuration. 
Two exposures of 1200 sec were taken at each fibre configuration. Fibres
unallocated to galaxy candidates were used for obtaining sky spectra, 
and wavelength calibration frames were taken for each fibre
configuration.
To maximise the efficiency of the available telescope time, we decided to
obtain spectra of all objects with APM magnitudes of $b_j\:\leq\:18.5$ in the
30 arcmin radius field of view.  Spectra were obtained of 69 objects using 3
fibre configurations.  These objects are listed in Table~\ref{gallist}, with
object designation given in column~1. The nomenclature of objects is
arbitrary, but the EXT designation indicates objects which were not selected
from the APM scans as galaxies, but were nevertheless added either because
their classification as galaxies were marginal or because of their proximity
to the QSO sightline. All these objects turned out to be stars.

Data were reduced using the {\tt wyffos} data reduction package (Lewis 1996)
running under IRAF. Redshifts of extracted spectra were measured by
cross-correlation with two radial velocity standards HD182572 ($v_r =
-100.5\pm 0.4$~\kms) and HD171391 ($v_r = +6.9\pm 0.2$~\kms) using the {\tt
rv} IRAF package. The resolution of our data, as measured from the calibration
arc lines, was 6.2~\AA , or 350~\kms\ at 5300~\AA ; cross-correlation of
galaxy spectra with template radial standards is known to give accuracies of
$\sim\:0.1\times$ FWHM, or $\sim\:35$~\kms\ for our data. Cross correlation of
the two radial velocity standards with each other easily found the correct
shift, confirming that the technique could certainly measure velocities to at
least 100~\kms .

Sky was subtracted from each spectrum using summed sky spectra obtained 
during object exposures.  Column 6 of Table~\ref{gallist} indicates the
quality of the final spectrum: `a' indicates a certain redshift, measured
independently from both of the two exposures of a given fibre configuration;
`b' also indicates a certain redshift (more than one feature is identified in
a spectrum) but the redshift could only be measured accurately from one of the
exposures; `c' means a redshift could be measured but the data were of low
quality and the redshift may be uncertain (this applied to only two objects in
our sample); `d' indicates that no redshift could be measured.

Of 69 objects observed, 63 redshifts were obtained. Of these, 44 were
galaxies.  Redshifts for 68 objects are given in column~5 of
Table~\ref{gallist} (Obj56 in the original list of 69 objects measured was
removed --- see \S2.1.2), along with the RA \& DEC of the object (cols. 2 and
3), the $b_j$ magnitude derived from the APM scans (col.~4), the angular
separation of the object on the plane of the sky from the QSO, $\theta$, in
arcmins, (col.~7), and the corresponding separation in units of \h, $\rho$,
when the redshift is known. Values of $\rho\:=\:0$ are
given for objects identified as stars.


\begin{table*}
\centering
\caption{Galaxy Redshifts within 30$'$ of Q1821+643}
\label{gallist}
\begin{tabular}{lcccrccr}
\hline
& RA
& Dec
&
&
&    
& $\theta$      
& \push{$\rho $}   \\
\push{Des}    
&\multicolumn{2}{c}{(J2000.0)} 
& $b_j$ 
& \push{$z_{\rm{gal}}$ }
& Qual. 
& ($'$)      
& \push{($h^{-1}$~kpc)}   \\
\hline
\multicolumn{8}{c}{WYFFOS data}\\
\hline
Ext    & 18:25:12.31 & +64:30:04.8  & 16.77 &$-$0.00040 & a  & 23.10 & 0.0 \\
Obj81  & 18:23:58.30 & +64:26:51.3  & 14.45 &0.12311  & a  & 14.50 & 1269.9 \\
Obj19  & 18:25:26.87 & +64:36:39.1  & 18.14 &$\ldots$ & e  & 27.71 &$\ldots$ \\
Obj54  & 18:24:40.36 & +64:34:54.0  & 17.00 &0.10642  & a  & 22.67 & 1761.9 \\
Obj39  & 18:25:04.56 & +64:40:35.8  & 18.32 &0.09583  & a  & 28.39 & 2020.8 \\
Obj58  & 18:24:33.84 & +64:38:51.2  & 14.82 &$\ldots$ & e  & 24.85 &$\ldots$ \\
Obj113 & 18:23:15.55 & +64:33:05.6  & 17.78 &0.19188  & b  & 15.08 & 1854.1 \\
Obj93  & 18:23:38.88 & +64:41:46.7  & 12.71 &0.08826  & a  & 23.83 & 1581.3 \\
Obj153 & 18:22:20.92 & +64:26:50.9  & 17.17 &0.19231  & a  & 6.75  & 831.2 \\
Obj144 & 18:22:31.70 & +64:35:28.4  & 12.90 &0.05047  & a  & 15.33 & 618.9 \\
Obj142 & 18:22:36.11 & +64:45:44.9  & 16.41 &$\ldots$ & e  & 25.49 &$\ldots$ \\
Obj167 & 18:22:07.61 & +64:39:32.4  & 17.47 &0.05039  & a  & 18.97 & 764.8 \\
Obj186 & 18:21:49.21 & +64:37:48.4  & 10.08 &0.05019  & a  & 17.22 & 691.7 \\
Obj198 & 18:21:36.70 & +64:45:22.6  & 17.09 &0.10725  & a  & 24.87 & 1945.4 \\
Obj216 & 18:21:21.38 & +64:49:40.4  & 18.08 &$\ldots$ & e  & 29.32 &$\ldots$ \\
Obj219 & 18:21:18.20 & +64:43:35.5  & 18.19 &$-$0.00008 & a& 23.37 & 0.0 \\
Obj264 & 18:20:14.22 & +64:38:56.5  & 16.50 &$-$0.00018 & a& 21.42 & 0.0 \\
Obj238 & 18:20:56.96 & +64:27:55.0  & 17.52 &0.07198  & a  & 9.78  & 543.5 \\
Obj293 & 18:19:40.64 & +64:30:38.5  & 17.92 &0.17955  & a  & 17.83 & 2089.2 \\
Obj318 & 18:18:53.04 & +64:30:37.4  & 18.47 &$-$0.00009 & a& 22.25 & 0.0 \\
Obj290 & 18:19:44.59 & +64:23:07.4  & 15.66 &0.08941  & a  & 14.56 & 977.0 \\
Obj334 & 18:18:14.83 & +64:18:22.4  & 16.95 &0.05106  & a  & 24.19 & 987.1 \\
Obj272 & 18:20:02.47 & +64:18:53.0  & 18.32 &0.02788  & a  & 12.54 & 290.5 \\
Obj332 & 18:18:17.71 & +64:16:15.9  & 17.41 &0.18002  & a  & 24.18 & 2838.7 \\
Obj338 & 18:18:02.25 & +64:14:04.2  & 16.91 &0.12103  & a  & 26.30 & 2271.7 \\
Obj327 & 18:18:35.45 & +64:14:20.8  & 16.49 &0.09538  & a  & 22.75 & 1612.9 \\
Obj245 & 18:20:48.75 & +64:18:52.4  & 18.29 &$-$0.00029 & a& 7.61  & 0.0 \\
Obj307 & 18:19:24.44 & +64:07:53.1  & 17.85 &0.17993  & a  & 20.91 & 2453.9 \\
Obj221 & 18:21:19.80 & +63:59:45.2  & 16.86 &$-$0.00015 & a& 21.25 & 0.0 \\
Obj146 & 18:22:30.82 & +64:04:12.9  & 15.41 &0.07171  & a  & 16.79 & 929.9 \\
Obj137 & 18:22:41.28 & +64:10:27.1  & 17.07 &$\ldots$ & e  & 11.23 &$\ldots$ \\
Obj112 & 18:23:16.01 & +64:04:52.9  & 14.11 &0.02754  & a  & 17.91 & 410.0 \\
Obj92  & 18:23:41.16 & +64:01:08.3  & 16.70 &0.07249  & a  & 22.52 & 1259.2 \\
Obj213 & 18:21:25.88 & +63:51:36.0  & 17.22 &0.00028  & a  & 29.21 & 0.0 \\
Obj204 & 18:21:37.94 & +63:51:24.4  & 18.01 &0.00070  & b  & 29.27 & 0.0 \\
Obj178 & 18:21:57.41 & +63:52:44.1  & 17.00 &0.00004  & a  & 27.87 & 0.0 \\
Obj165 & 18:22:09.95 & +64:09:36.9  & 17.87 &0.08410  & a  & 11.08 & 705.3 \\
Obj151 & 18:22:25.74 & +63:53:51.5  & 18.46 &0.00005  & a  & 26.93 & 0.0 \\
Obj105 & 18:23:25.41 & +64:08:34.0   & 9.38 &0.05033  & a  & 15.39 & 619.8 \\
Obj53  & 18:24:40.55 & +64:02:45.1  & 17.09 &0.09681  & a  & 25.20 & 1809.2 \\
Obj67  & 18:24:21.54 & +64:06:26.2   & 7.30 &0.05106  & a  & 21.14 & 862.6 \\
Obj34  & 18:25:08.16 & +64:09:17.3  & 18.37 &0.13853  & a  & 23.63 & 2273.5 \\
Obj15  & 18:25:46.41 & +64:07:56.8  & 18.22 &0.07200  & b  & 27.94 & 1553.0 \\
Obj63  & 18:24:28.05 & +64:16:41.5  & 17.98 &0.08470  & c  & 16.81 & 1076.7 \\
Obj172 & 18:22:02.73 & +64:21:38.7  & 17.60 &0.12160  & a  & 1.20  & 104.0 \\
Obj217 & 18:21:22.44 & +64:28:28.7  & 18.08 &0.12245  & a  & 8.72  & 760.4 \\
Obj180 & 18:21:53.90 & +64:19:49.8  & 18.34 &$-$0.00012 & a& 0.85  & 0.0 \\
Obj345 & 18:17:33.14 & +64:19:29.9  & 17.38 &0.00000  & b  & 28.61 & 0.0 \\
Obj271 & 18:20:04.95 & +64:19:44.3  & 18.47 &$-$0.00014 & a& 12.18 & 0.0 \\
Obj242 & 18:20:53.45 & +64:19:37.1  & 18.10 &0.11155  & b  & 6.97  & 563.2 \\
Obj212 & 18:21:25.61 & +64:02:11.0  & 17.79 &0.19406  & a  & 18.74 & 2322.7 \\
Obj192 & 18:21:41.26 & +63:51:37.2  & 15.81 &0.02380  & b  & 29.04 & 578.4 \\
Obj132 & 18:22:47.01 & +63:59:28.6  & 15.49 &0.07201  & a  & 21.82 & 1213.0 \\
Obj140 & 18:22:36.32 & +64:12:42.1  & 17.96 &$-$0.00034 & a& 8.97  & 0.0 \\
Obj57  & 18:24:33.26 & +64:16:13.9  & 16.63 &0.05305  & a  & 17.48 & 738.6 \\
Obj72  & 18:24:12.74 & +64:24:40.1  & 18.02 &$-$0.00011 & a& 15.21 & 0.0 \\
Obj97  & 18:23:34.34 & +64:18:35.1  & 17.01 &0.05099  & a  & 10.72 & 436.9 \\
Ext    & 18:24:09.55 & +64:19:33.7  & 18.17 &$-$0.00035 & a& 14.37 & 0.0 \\
\end{tabular}
\end{table*}
\setcounter{table}{0}
\begin{table*}
\centering
\caption{Cont.}
\begin{tabular}[t]{lcccrccr}
& RA
& Dec
&       
&          
&    
& $\theta$       & \push{$\rho $}   \\
\push{Des}    
&\multicolumn{2}{c}{(J2000.0)} 
& $b_j$ 
& \push{$z_{\rm{gal}}$ }
& Qual. 
& ($'$)      
& \push{($h^{-1}$~kpc)}   \\
\hline
Obj10  & 18:26:01.05 & +64:23:16.4  & 16.17 &0.07623  & a  & 26.51 & 1549.3 \\
Obj73  & 18:24:12.45 & +64:35:12.7  & 18.24 &$-$0.00020 & a& 20.64 & 0.0 \\
Obj90  & 18:23:42.40 & +64:32:53.3  & 16.81 &0.05242  & a  & 16.72 & 698.8 \\
Obj87  & 18:23:44.90 & +64:42:45.9  & 18.06 &0.08849  & a  & 25.00 & 1662.7 \\
Obj148 & 18:22:29.27 & +64:41:05.4  & 16.76 &0.08807  & a  & 20.77 & 1375.7 \\
Obj347 & 18:17:26.18 & +64:19:43.7  & 18.24 &$\ldots$ & e  & 29.35 &$\ldots$ \\
Obj230 & 18:21:08.73 & +63:52:09.0  & 18.04 &0.15039  & a  & 28.94 & 2968.2 \\
Obj118 & 18:23:09.33 & +63:52:23.4  & 17.14 &0.00000  & a  & 29.29 & 0.0 \\
Obj138 & 18:22:40.55 & +64:09:43.7  & 18.42 &0.12050  & a  & 11.85 & 1019.9 \\
Obj130 & 18:22:50.78 & +64:09:40.8  & 18.40 &0.00000  & b  & 12.38 & 0.0 \\
\hline
\multicolumn{8}{c}{HYDRA data}\\
\hline
Ext    & 18:23:57.96 & +64:11:07.7  &  8.61 &$-$0.00014 & a& 16.18 & 0.0 \\
Obj131 & 18:22:50.92 & +64:09:19.8  & 12.71 &0.05086  & a  & 12.70 & 516.4 \\
Obj56  & 18:24:34.11 & +64:19:04.1  & 14.80 &0.05050  & a  & 17.07 & 689.5 \\
Obj58  & 18:24:33.84 & +64:38:51.2  & 14.83 &0.09605  & b  & 24.85 & 1772.2 \\
Obj181 & 18:21:52.77 & +64:38:59.5  & 15.32 &0.05050  & a  & 18.39 & 742.9 \\
Obj142 & 18:22:36.11 & +64:45:44.9  & 16.42 &0.04970  & a  & 25.49 & 1014.7 \\
Ext    & 18:25:12.31 & +64:30:04.8  & 16.77 &$-$0.00030 & a& 23.10 & 0.0 \\
Obj216 & 18:21:21.38 & +64:49:40.4  & 18.08 &0.05683  & c  & 29.32 & 1318.9 \\
Obj19  & 18:25:26.87 & +64:36:39.1  & 18.14 &0.09517  & a  & 27.71 & 1960.9 \\
Ext    & 18:24:09.55 & +64:19:33.7  & 18.18 &$-$0.00025 & a& 14.37 & 0.0 \\
Obj347 & 18:17:26.18 & +64:19:43.7  & 18.25 &0.08210  & b  & 29.35 & 1829.8 \\
Obj62  & 18:24:29.09 & +64:01:05.3  & 18.57 &0.08420  & a  & 25.58 & 1630.0 \\
Obj13  & 18:25:53.58 & +64:19:18.5  & 18.61 &0.11159  & b  & 25.63 & 2071.7 \\
Obj309 & 18:19:23.46 & +63:59:26.8  & 18.61 &0.00005  & a  & 26.98 & 0.0 \\
Obj331 & 18:18:17.57 & +64:34:37.2  & 18.65 &0.07836  & a  & 27.51 & 1647.0 \\
Obj20  & 18:25:20.57 & +64:16:35.7  & 18.67 &0.09697  & a  & 22.41 & 1611.2 \\
Obj313 & 18:19:07.28 & +64:36:50.3  & 18.67 &$\ldots$ & e  & 24.46 &$\ldots$ \\
Obj269 & 18:20:08.55 & +64:00:06.1  & 18.71 &0.00006  & a  & 23.67 & 0.0 \\
Obj241 & 18:20:57.82 & +63:56:00.8  & 18.71 &0.00040  & a  & 25.43 & 0.0 \\
Ext5   & 18:22:11.40 & +64:28:43.0  & 18.77 &$-$0.00013 & a& 8.26  & 0.0 \\
Obj234 & 18:21:00.73 & +64:37:54.3  & 18.77 &0.26650  & c  & 18.34 & 2813.8 \\
Obj314 & 18:19:10.44 & +64:07:27.4  & 18.77 &0.09458  & a  & 22.39 & 1576.1 \\
Obj297 & 18:19:34.62 & +64:42:21.0  & 18.77 &0.16410  & a  & 26.60 & 2915.55\\
Obj328 & 18:18:28.71 & +64:38:03.6  & 18.80 &0.16550  & a  & 28.43 & 3136.0 \\
Obj205 & 18:21:35.27 & +64:25:24.0  & 18.83 &0.18900  & a  & 5.35  & 650.7 \\
Obj38  & 18:25:04.35 & +64:25:23.1  & 18.85 &0.12005  & a  & 20.79 & 1784.0 \\
Obj275 & 18:19:56.14 & +64:40:20.0  & 18.85 &0.00008  & a  & 23.64 & 0.0 \\
\hline
\end{tabular}
\end{table*}

\subsubsection{HYDRA data}

Additional spectra of objects in the field of Q1821+643 were obtained with the
HYDRA instrument on the WIYN telescope at Kitt Peak.
HYDRA is similar in design to WYFFOS in that a fibre positioner
feeds optical fibres to a bench spectrograph.  The field of view (30 arcmin
radius) is also the same. Observations were made as part of the WIYN Queue
Observing Experiment on September 3rd 1996 using 316 lines~mm$^{-1}$ grating
centered at 5800~\AA\ and a thinned 2048x2048 CCD covering a wavelength
range of $\approx 4000$~\AA .

Three exposures of 2400 sec each were taken with one fibre configuration.
Fibres unallocated to galaxy candidates were again used for obtaining sky
spectra, and wavelength calibration frames were taken before and after the
configuration exposures.  Data were reduced using the {\tt dohydra} data
reduction package (Valdes 1992) running under IRAF. Redshifts of 
spectra were measured by cross-correlation with the same two radial velocity
standards used for the WYFFOS data. The resolution as measured from the
comparison arc lines was 8.4~\AA , or 430~\kms\ at 5800~\AA .

Spectra were obtained of 80 objects. 58 of these had already been
observed with WYFFOS.  Several HYDRA spectra yielded redshifts which could
not be be measured in the WYFFOS data. These objects, as well as those not
covered by the WYFFOS configurations, are listed separately in
Table~\ref{gallist}. For objects with redshifts identified from both sets of
data, only one galaxy was found to have a discrepant redshift. This galaxy
(Obj56 in Table~\ref{gallist}) was one of the two spectra with a `c' quality
flag in the WYFFOS data, and we adopt the HYDRA measurement of its redshift in
Table~1.

\subsubsection{Completeness}

\begin{figure*}
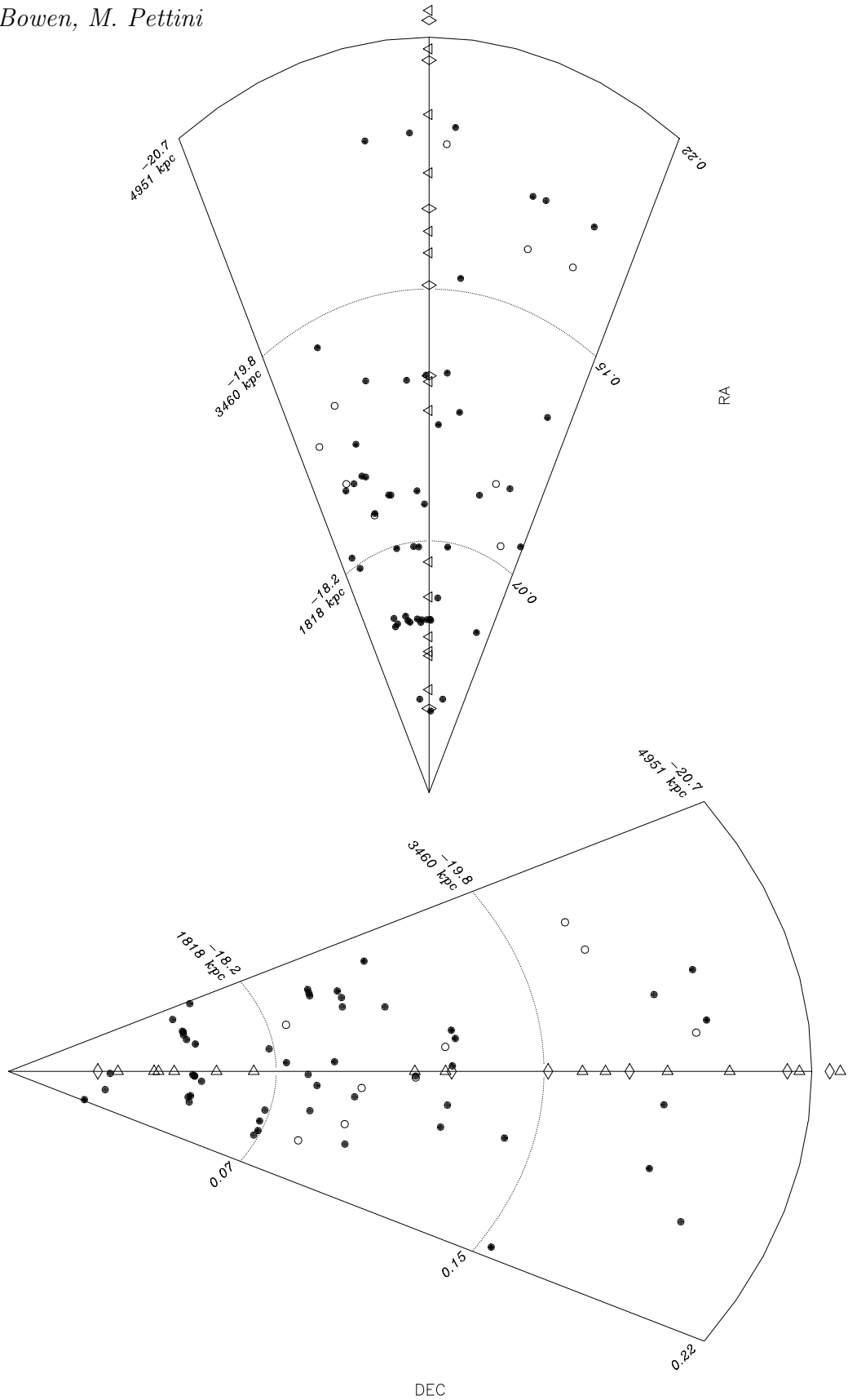

\vspace*{-2cm}
\centerline{
\psfig{figure=wedge1.ps,height=15cm,angle=0}}
\vspace{-2cm}
\centerline{
\psfig{figure=wedge2.ps,height=11cm,angle=270}}
\caption{Wedge diagrams showing the spatial distribution of galaxies observed
with WYFFOS and HYDRA as a function of redshift, plotted in RA (top) and DEC
(bottom). Redshifts of \lya\ lines towards Q1821+643 are plotted along the
central axis; diamonds indicate lines with $W_r\: \geq\: 0.2$~\AA , 
triangles $W_r \:<\:0.2$~\AA. Galaxies forming part of our magnitude limited
sample of $b_j\:\leq\:18.5$ are shown as filled circles, fainter galaxies are
open circles. The absolute
magnitudes of this limit at the redshifts marked along the right axis of each
cone are
shown on the left side, along with the comoving angular distance that 30
arcmin corresponds to at the same redshift.}
\label{fig:wedges}
\end{figure*}

As noted above, our initial goal was to observe all galaxies within 30 arcmins
of the sightline of Q1821+643 down to a magnitude of $b_j\:=\:18.5$.
Combining the WYFFOS and HYDRA data, we measured 68 of 71 galaxy candidates, a
completeness of 96\%.  17 of 68 candidates, or 25\%, were found to be stars. A
large number of stellar spectra were expected, since we included objects
marginally identified as galaxies to ensure that compact galaxies were not
exlcuded in our sample.  The distribution of confirmed galaxies in Right
Ascension and Declination vs.\ redshift is shown in Fig.~\ref{fig:wedges}.  We
believe the errors in the APM magnitudes may be within $\pm 0.5$ mags, but CCD
images taken over the entire field of view of the spectrographs are needed to
confirm this figure. Thus, the total galaxy sample considered here consists of
51 galaxies brighter than $b_j\:=\:18.5$ within 30 arcmin of Q1821+643 on the
plane of the sky.

\subsection{Identification of \lya\ systems}

There exist several independent observations of Q1821+643 made with \hst ,
covering different wavelength ranges and obtained at different resolutions.
The first data were obtained with the {\it FOS} by Bahcall~\etal\ (1992,
hereafter B92) using the G130H, G190H, and G270H gratings, and subsequently
reanalysed as part of the Absorption Line Key Project (Bahcall~\etal\ 1993,
hereafter KPI). Higher resolution data from the GHRS were obtained by Savage,
Sembach \& Lu (1995, hereafter SSL) with the G160M grating over two small
wavelength regions from $1232-1269$~\AA\ and $1521-1558$~\AA . Finally, GHRS
 G140L data are available from the \hst\ Archive, taken  30-March-1996 and
covering the wavelength range $\simeq\:1255-1530$~\AA .

Comparing the list of \lya\ absorption lines identified in FOS spectra by B92
and KPI shows several inconsistencies which arise from changes in their method
of line identification. Further, the high resolution GHRS spectra obtained by
SSL show that some of the initial identifications in the FOS spectra were
erroneous. To derive a list of \lya\ absorption systems to compare with
galaxies detected in our spectroscopic survey, we have extracted the FOS and
GHRS data from the \hst\ Archive and reanalysed the data with our own software.
Our identification of absorption lines follows similar procedures to those
described by Young~\etal\ (1979) and by B92. (These procedures differ from
those used subsequently by KPI who used the instrumental Line Spread Function
(LSF) of the spectrograph to weight the summation of pixels comprising a line
in the determination of its equivalent width.)

There are several differences in our analysis to those used by B92 and
KPI. First, we have corrected the FOS wavelength scales by applying a shift to
match the rest wavelengths of low ionization lines arising from Galactic
absorption in the spectra to the velocity of the bulk of H~I emission measured
at 21~cm. We take the heliocentric velocity of this gas to be at $\simeq\:
-40$~\kms\ (Lockman \& Savage 1995). Second, no accurate LSFs are available
for the FOS aperture used ($0\farcs25\times 2\farcs0$) so we have not used an
LSF to estimate the equivalent widths of lines. Instead, we have taken the
number of pixels over which an absorption feature is detected to be 11.  This,
for example, is the extent of the LSF for the $1.''0$ aperture (see BBP).
Third, error arrays supplied by the pipeline wavelength calibration of FOS
spectra do not include any contribution from the background or from other
sources such as scattered light, and are therefore inadequate for determining
the reality of a feature in a spectrum.  We have therefore constructed arrays
from measured rms deviations in regions of a spectrum clearly free of
absorption lines, and used these for our error arrays. The flux arrays used to
measure the noise are those prior to the data being resampled to a linear
wavelength scale because resampling introduces a smoothing of the flux.  For
GHRS data, the pipeline calibrated error arrays are true representations of
the noise in the data, and can be used for detection and measurement of
absorption lines.

Our final line lists consist of lines at wavelength $\lambda_{\rm{obs}}$, with
observed equivalent widths, $W$, which are $\geq 3\sigma(W)$, where
$\sigma(W)$ is the error in the equivalent width at $\lambda_{\rm{obs}}$.  For
lines which are in common with those found by B92 and KPI, the majority of our
measurements of $W$ are within $\pm 3\sigma(W)$ of their measurements. Hence
our measurements of $W$ give values close to those already published. Only if
a line sits on the peak of an emission line do equivalent widths differ by
more that $\pm 3\sigma(W)$, because errors inherent in continuum fitting
around the emission line mean that the true value of $\sigma (W)$ is greater
than that estimated from the noise. Of the 18 lines we detect in FOS or
published GHRS data, 15 have been identified by B92, SSL, and/or KPI. We
detect an additional \lya\ line in the GHRS data at 1252.1~\AA , which is
weak, but present at the 3$\sigma$ level in our analysis. Another two
lines are marginal detections in the FOS G130H data, but are confirmed in the
unpublished GHRS G140L spectrum.

The \lya\ lines we identify are listed in Table~\ref{linelist}.  Column 4
lists the absorption redshift, $z_{\rm{abs}}$, of the absorption line; column
5, $z_{\rm{GHRS}}$, indicates whether the redshift has been determined from
the high resolution G160M GHRS spectra of SSL, either by direct measurement of
the \lya\ line or higher order Lyman lines of the same absorption system.
Measurements of $z_{\rm{abs}}$ using these spectra are more accurate than
those derived from the FOS data.  Column 6 lists the absolute magnitude we
reach in our survey at $z_{\rm{abs}}$ given the limiting observed magnitude of
$b_j\:=18.5$.  Column 7 lists in which spectra the lines have been identified,
while in column 8 we record whether the authors mentioned above also
identified the line.

\begin{table*}
\centering
\begin{minipage}{110mm}
\caption{\lya\ systems along the line of sight to Q1821+643}
\label{linelist}
\begin{tabular}{llclccrl}
$\lambda_{\rm{obs}}$
& $W$
& $\sigma (W)$
& 
& 
&
&
& \\
(\AA )
& (\AA )
& (\AA )
& $z_{\rm{abs}}$  
& $z_{\rm{GHRS}}$
& $M_{\rm{lim}}$
& Instr.\footnote{Line detected in --- (1) GHRS G160M spectrum; 
(2) FOS G130H spectrum;
(3) GHRS G140L spectrum} 
& Who else?\footnote{Line also 
found by: SSL---Savage, Sembach \& Lu 1995; B92---Bahcall~\etal\ 1992;
KPI---Bahcall~\etal\ 1993}\\
\hline
1245.33  & 0.32 & 0.02 & 0.0245 & y   & $-15.8$ & 1 	& SSL \\
1252.10  & 0.05 & 0.02 & 0.0300 & y   & $-16.3$ & 1	&     \\
1264.17  & 0.19 & 0.02 & 0.0399 & y   & $-16.9$ & 1,3 	& SSL \\
1265.59  & 0.04 & 0.01 & 0.0411 & y   & $-16.9$ & 1 	& SSL \\
1270.9   & 0.13 & 0.01 & 0.0454 & n   & $-17.2$ & 2,3 	&     \\
1285.0   & 0.06 & 0.01 & 0.0570 & n   & $-17.7$ & 2,3 	&     \\
1297.3   & 0.13 & 0.01 & 0.0672 & n   & $-18.1$ & 3 	& KPI \\
1351.0   & 0.10 & 0.01 & 0.1113 & n   & $-19.2$ & 3 	&     \\
1361.2   & 0.05 & 0.01 & 0.1197 & n   & $-19.3$ & 3     &	\\
1363.2   & 0.71 & 0.06 & 0.1214 & n   & $-19.4$ & 2,3 	& B92,KPI \\
1395.3   & 0.35:& 0.06 & 0.1478 & n   & $-19.8$ & 2,3   & B92,KPI \\
1406.8   & 0.06 & 0.01 & 0.1572 & n   & $-19.9$ & 3 	&     \\
1414.4   & 0.07 & 0.01 & 0.1635 & n   & $-20.0$ & 3 	&     \\
1422.5   & 0.62 & 0.06 & 0.1701 & n   & $-20.1$ & 2,3 	& B92,KPI\\
1435.1   & 0.23 & 0.06 & 0.1805 & n   & $-20.3$ & 2,3 	& KPI \\
1455.8   & 0.13 & 0.01 & 0.1975 & n   & $-20.5$ & 3 	&     \\
1474.8   & 0.54 & 0.07 & 0.2133 & y   & $-20.6$ & 2,3	& B92,KPI \\
1478.7   & 0.20 & 0.07 & 0.2166 & y   & $-20.7$ & 2,3	& KPI \\
1489.3   & 1.01 & 0.07 & 0.2249 & y   & $-20.8$ & 2,3 	& B92,KPI \\
1492.6   & 0.16:& 0.02 & 0.2278 & n   & $-20.8$ & 2,3  	& B92 \\ 
1513.7   & 0.06 & 0.02 & 0.2452 & n   & $-21.0$ & 3 	&             \\
1529.46  & 0.21 & 0.02 & 0.2581 & y   & $-21.1$ & 1,2,3 & SSL,B92     \\
1533.63  & 0.22 & 0.02 & 0.2616 & y   & $-21.1$ & 1,2	& SSL,B92,KPI \\
1539.74  & 0.24 & 0.02 & 0.2666 & y   & $-21.1$ & 1,2 	& SSL,B92,KPI \\
1576.6   & 0.56 & 0.03 & 0.2967 & y   & $-21.4$ & 2 	& B92,KPI     \\
\hline
\end{tabular}
\end{minipage}
\end{table*}

We identify lines at 1264.2 and 1270.9~\AA\ as \lya\ absorption at redshifts
of $z_{\rm{abs}}\:=\:0.0399$ and $0.0454$. However, it is also possible that
these are actually an O~VI absorption doublet at $z_{\rm{abs}}\:=\:0.2249$,
from an absorption system identified by its strong \lya\ and Ly$\beta$
absorption (Tripp~\etal\ 1997). We maintain that the lines are indeed
low-redshift \lya\ absorption for two reasons: a) the
$z_{\rm{abs}}\:=\:0.2249$ Ly$\beta$ and O~VI$\lambda 1031$ lines do not match
up in velocity (although this could be because of additional lower column
density gas which shows up in H~I absorption but not O~VI); and b) there is no
absorption from C~IV, Si~IV or N~V (or indeed from any other metal species) at
$z_{\rm{abs}}\:=\:0.2249$. Clearly, higher sensitivity observations are needed
to test the identification of these lines further.

\begin{figure*}
\centerline{\psfig
{figure=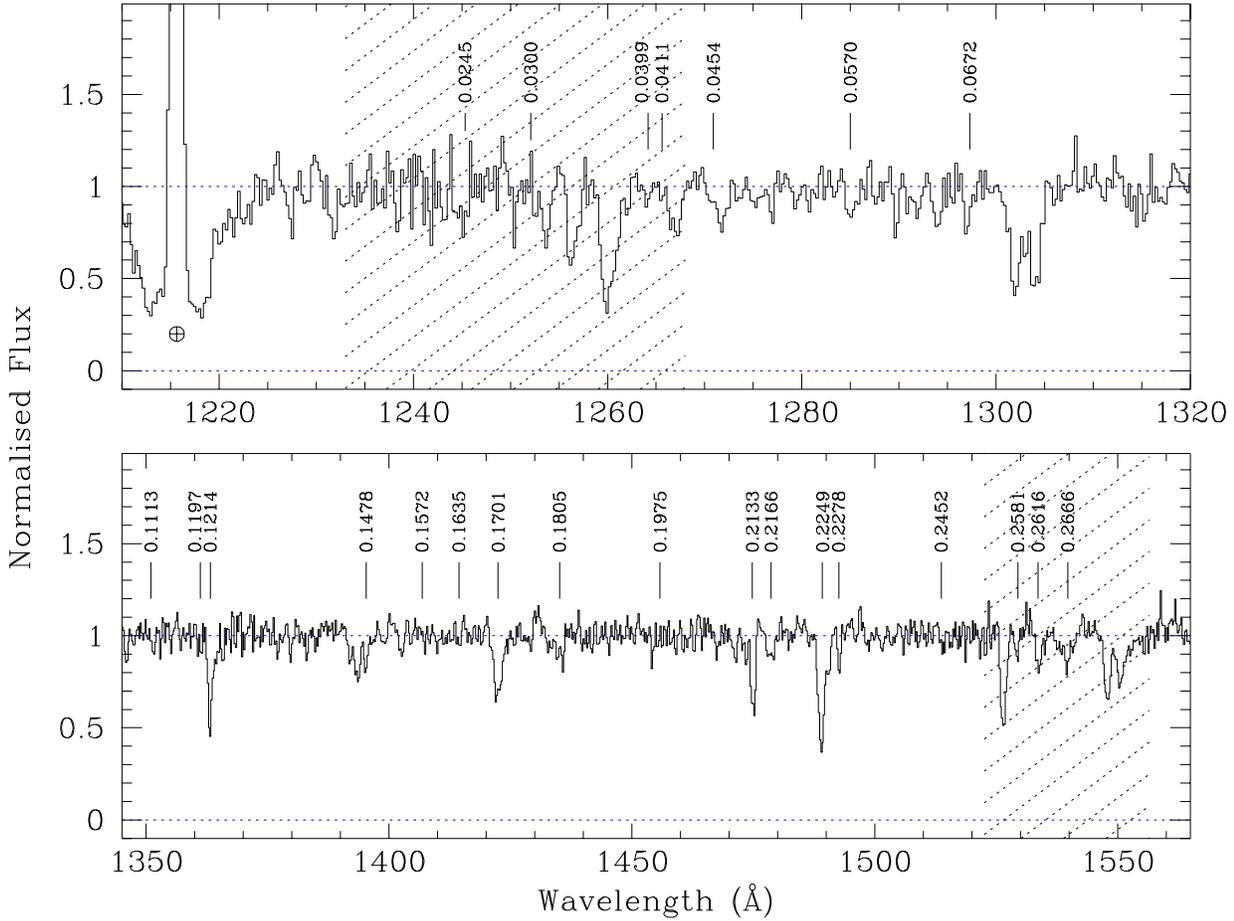,height=14cm,angle=270}}
\caption{Portions of the normalised spectra 
of the FOS G130H spectrum of Q1821+643
originally presented by B92 \& KPI. All the \lya\ lines listed in Table 2 are
indicated. The shaded regions represent wavelength ranges spanned by the
high resolution GHRS G160M data obtained by SSL; tick marks which do not
appear to correspond to any obvious lines 
here indicate the positions of weaker \lya\ absorption revealed by 
GHRS G160M data.
Similarly, lines
marked between $\sim\:1270-1320$~\AA\ which appear marginal here
can be
seen in the GHRS G140L spectrum extracted from the \hst\ Archive (Fig.~3).}
\label{fig:g130} 
\end{figure*}

\begin{figure*}
\vspace*{-5cm}
\centerline{\psfig
{figure=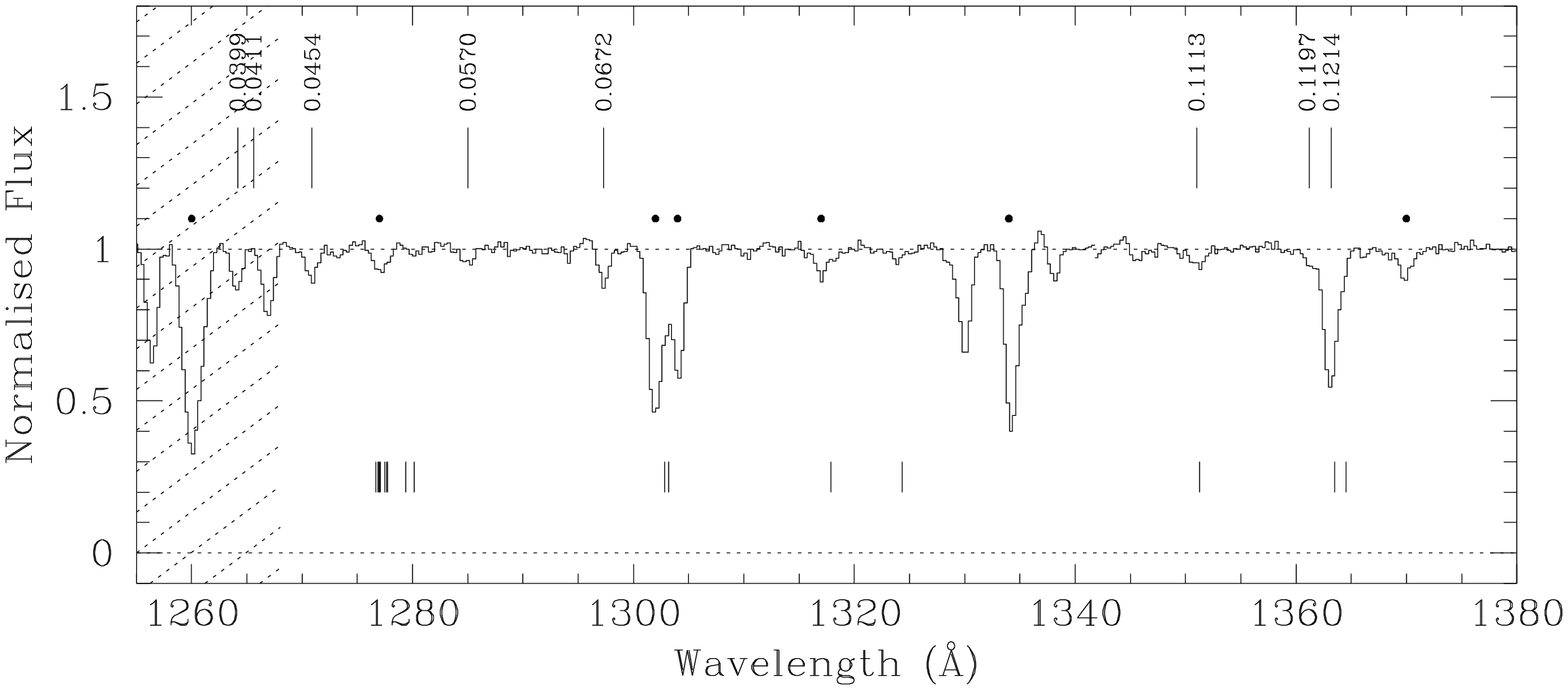,height=15cm,angle=270}}
\caption{Portions of the normalised G140L spectrum 
extracted from the \hst\ Archive, with the positions of \lya\ lines marked. 
 Data are binned at the original dispersion
of 0.29~\AA~pix$^{-1}$. 
See caption to Fig.~2 for significance of shaded
region. Redshifts of identified \lya\ lines are shown above the spectrum,
along with the positions of absorption lines from the Milky Way indicated by
$\bullet$; the expected wavelengths of \lya\ absorption from all the galaxies
listed in Table~1 which are within $\rho \:\leq\:1$~\mpc\ are shown by tick
marks below the spectrum.}
\label{fig:g140} 
\end{figure*}

In Figs.~\ref{fig:g130} and~\ref{fig:g140} we show portions of the FOS
G130H and GHRS G140L normalised spectra and \lya\ systems identified.
Comparisons between features found in both FOS G130H and GHRS G140L spectra
suggest that the values of $z_{\rm{abs}}$ in Table~\ref{linelist} are
consistent to within $|\Delta z| \:\approx\: 0.0002$.

\subsection{NED galaxies}

Due to the pencil-beam nature of our survey, the number of galaxies sampled
below $z< 0.05$ begins to decline as the 1 degree angular size of the
sky samples smaller physical radii, and hence volume. We have therefore
supplemented our list of galaxies by searching the NASA Extragalactic
Database (NED) for all galaxies with known redshifts within 300 arcmins of
the QSO line of sight. Of 136 galaxies found, none were within 500~\h\ of
the QSO sightline.
A wider search for galaxy clusters over 30 degrees revealed only two clusters
close to the QSO sightline. ZwCl~1638.4+6038 at $z=0.01634$ is 10.3~\mpc\
away, while Abell~2295 at $z=0.05080$, is 13.0~\mpc\ from the line of
sight. The 13 galaxies we found at $z\:\simeq\:0.051\pm0.02$ in 
Fig.~\ref{fig:wedges} are presumably
members of a larger structure of which Abell~2295 is part.

\section{Discussion}

Since the aim of our programme is to study the association of \lya\ lines with
low-redshift galaxies, we have detected very few galaxies beyond $z\:=\:0.2$
(see Fig.~\ref{fig:wedges}). The magnitude limit of our survey
($b_j\:\leq\:18.5$) means we 
become insensitive to galaxies fainter than $L^*$ beyond 
$z\:\approx\:0.1$. Deeper searches for galaxies close to the QSO line of 
sight include those by Schneider~\etal\ (1992), who have
detected several galaxies within one arcmin.
Although most appear to be at similar redshifts to the QSO itself, one galaxy
(labelled G in their Table~1, 73~\h\ from the QSO sightline) is within
600~\kms\ of the $z_{\rm{abs}}\:=\:0.2249$ and 0.2278 absorption
systems. Also, Le Brun~\etal\ (1996) have detected a galaxy (their \#8) at a 
redshift of $0.1703$, 266~\h\ from the QSO line of sight, and close to 
the redshift of the $z_{\rm{abs}}\:=\:0.1701$ system listed in Table~2.
However, since these galaxies have been detected in surveys with 
different magnitude limits, survey completion, and extent on the sky than 
our own, we do not include them in our subsequent analysis. 

\subsection{\lya -absorbing cross-section of individual galaxies}

We begin first by considering the limits we can set on the size of individual
galaxy halos from our data. Table~1 contains 49 galaxies within 2~\mpc\ of the
QSO line of sight.  For each galaxy at a separation $\rho $ from the QSO
sightline we have searched for \lya\ absorption in the QSO spectra within $\pm
600$~\kms\ of the galaxy's measured redshift. 18 of these galaxies could not
be searched for absorption, however, since any \lya\ absorption which did
arise at the redshift of the galaxy would be obscured by Milky Way absorption
lines or other lines from higher redshift systems. Also, six galaxies would
have \lya\ absorption lines blended with the wings of weak ($W<\:0.2$~\AA )
Galactic lines; in these cases it has only been possible to search for \lya\
absorption from the galaxy within $\pm(200-300)\mp600$~\kms\ of its velocity.
Table~\ref{neargals} lists the first 13 galaxies within 1~\mpc\ for which we
were able to search for absorption, and the detection of any \lya\
lines. $\Delta v$ given in column 7 is the velocity difference derived from
$z_{\rm{gal}}-z_{\rm{abs}}$.

\begin{table*}
\centering
\begin{minipage}{100mm}
\caption{Galaxies with $\rho\:\leq\:1$~\mpc\ from the sightline of  Q1821+643}
\label{neargals}
\begin{tabular}{llcccrr}
    &                &                &$\rho$  & & $W_r$ & $\Delta v$ \\
Des & $z_{\rm{gal}}$ & $M_{\rm{gal}}$ & (\h )
& $z_{\rm{abs}}$
& (\AA )
& (\kms )\\ \hline
Obj172 & 0.12160 & $-20.3$ & 104.0 & 0.1214 & 0.63       & $54$ \\
Obj272 & 0.02788 & $-16.3$ & 290.5 & ...    & $<0.05$    & ... \\
Obj112 & 0.02754 & $-20.5$ & 410.1 & ...    & $<0.05$    & ... \\
Obj242 & 0.11155  & $-19.6$ & 563.2 & 0.1113 & $0.09$     & 67  \\
Obj192 & 0.02380  & $-18.5$ & 578.4 & 0.0245 & $0.31$     & $-205$ \\
Obj205 & 0.18900  & $-20.0$ & 650.7 & ...    & $<0.03$	 &... \\
Obj90  & 0.05242  & $-19.2$ & 698.8 & ...    & $<0.03$	 &... \\
Obj165 & 0.08410  & $-19.2$ & 705.3 & ...    & $<0.05$	 &... \\
Obj57  & 0.05305  & $-19.4$ & 738.6 & ...    & $<0.03$	 &... \\
Obj153 & 0.19231  & $-21.7$ & 831.2 & ...    & $<0.03$	 &... \\
Obj67  & 0.05106  & $-22: $ & 862.6 & ...    & $<0.10$	 &... \\
Obj290 & 0.08941  & $-21.5$ & 977.0 & ...    & $<0.03$	 &... \\
Obj334 & 0.05106  & $-19.0$ & 987.1 & ...    & $<0.10$	 &... \\
\hline
\end{tabular}
\end{minipage}
\end{table*}

\begin{figure}
\centerline{\psfig
{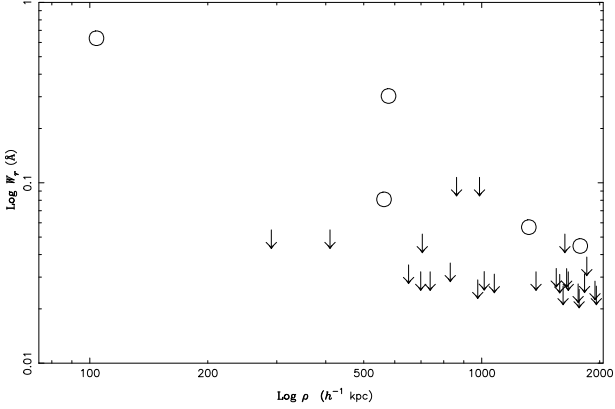}}
\caption{Plot of the distribution of  $\rho $ {\it vs.}\ rest equivalent
width, $W_r$, for 
galaxies in our survey. Open circles represent detections of \lya\ lines
within $\Delta v\:=\:\pm600$~\kms\ of a galaxy's velocity.
}
\label{fig:percent} 
\end{figure}

Fig.~\ref{fig:percent} plots $\rho $ vs.\ $W_r$ for galaxies where \lya\ could
be searched for; circles represent detections of \lya\ absorption, limits are
given by arrows.  We find three galaxies within 500~\h\ of the QSO line of
sight, and detect \lya\ absorption from the galaxy closest to the line of
sight (Obj172) at $z_{\rm{gal}}=0.1216$, $\rho \:=\: 104$~\h .  The absorption
line is moderately strong, with a rest equivalent width of $W_r = 0.63$~\AA ,
and there is no associated C~IV absorption to $3\sigma$ equivalent width
limits of $0.12$~\AA .  The galaxy shows strong [O~II], [O~III], and H$\beta$
emission, as well as strong Balmer lines, indicating recent star formation.
However, the absorbing galaxies found by LBTW generally did not show emission
lines, and it is unlikely that any correlation exists between the star
formation implied by the spectral features in Obj172's spectrum and the
existence of an extended gaseous halo.

The next closest galaxies are at impact parameters of 291 and 410~\h ; they
show no absorption to extremely low equivalent width limits of $0.05$~\AA\
(their redshifts were covered by SSL's high resolution GHRS data; these data
also show that no local Galactic lines prevent detection of \lya\ at the
galaxy redshifts).  One other galaxy is within $0.5$~\mpc, (Obj97 at
$z_{\rm{gal}}=0.05099$, $\rho \:=\:437$~\h ) but is not included in
Fig.~\ref{fig:percent} or Table~3 because the redshift at which \lya\ is
expected is obscured by a weak line at 1277.1~\AA\ which is probably
Galactic C~I. The line could in principle be \lya\ at
$z_{\rm{abs}}\:=\:0.0505$, since there is no corresponding C~I$\lambda
1280$. However, C~I$\lambda 1277$ is detected in higher resolution GHRS data
at roughly the same strength as observed here, and a simple Voigt profile fit
to the C~I$\lambda 1277$ line suggests that C~I$\lambda 1280$ would be not be
detected in spectra of the sensitivity of the G140L data. This is unfortunate,
because the C~I$\lambda 1277$ line obscures any \lya\ absorption from the
group of galaxies found at $z\:\sim\:0.05$.

The detection of a single absorbing galaxy 104~\h\ from the QSO line of sight
is consistent with LBTW's conclusion that galaxies are surrounded by spherical
halos of 160~\h\ radius.  There is good evidence from LBTW and BBP that
individual galaxies are not surrounded by halos with radii greater than
$\sim\:200-300$~\h , and although our data do not well sample galaxy halos on
sizes $\rho\:=\:100-500$~\h , the data continue the trend of non-detections
beyond $\rho\:=\:500$~\h , confirming that the majority of galaxies do not
possess \lya\ absorbing halos beyond 500~\h . There are, however, a few cases
where \lya\ lines arise at the same redshifts as galaxies. This is similar to
what BBP found on scales of $100-500$~\h , where a few absorber-galaxy
associations were found despite the general trend of non-detections. In that
work, it was impossible to test whether those associations were statistically
significant or not. Here, however, we have a controlled sample of galaxies and
absorption systems which allows us to test whether the detection of
\lya\ at the same redshifts as galaxies is
coincidental, although with our data we are mainly 
testing associations on scales of
$>\:200$~\h .

Another important point to emphasise is that the majority of \lya\ lines lines
listed in Table~2 are weak, $W_r\:<\:0.2$.  These lines are not detectable in
the FOS spectra of Q1821+643 [Bahcall~\etal\ (1996) in their analysis of FOS
Key Project data set a canonical value of 0.24~\AA\ as the limit to which FOS
spectra are sensitive over a large spectral range], and are only identified
clearly in the more sensitive GHRS data.  This makes it even more important to
test whether apparent galaxy-absorber associations arise purely from chance,
because the density of weak lines with redshift is very much higher than
strong ones, increasing the chances of `accidental' coincidence close to the
line of sight.  The statistical significance of all associations shown in
Fig.~4 are discussed in detail in \S3.3.

\subsection{Association of \lya -clouds with galaxies}

In Table~\ref{nearlines} we list the ten \lya\ lines for which we could have
detected a galaxy with luminosity brighter than $\approx\:L^*$, and any
galaxies found within $\pm 600$~\kms\ of the absorption line. The limit to
$\rho $ when no galaxy is found depends on the distance that 30 arcmins
corresponds to at the redshift of a \lya\ line; these lower limits are given
in column 6, and range from $\simeq\:0.75-1.6$~\mpc .  Column 7 gives the
proper separation, $s$~\mpc , between absorber and galaxy assuming that
$\Delta v$ represents cosmological (i.e. ``Hubble Flow'') distances.
$M_{\rm{lim}}$ is again listed in column 8 to highlight the limit to which we
are able to detect galaxies.

\begin{table*}
\centering
\begin{minipage}{100mm}
\caption{Nearest galaxies to \lya\ lines within 600~\kms }
\label{nearlines}
\begin{tabular}{cclcrccc}
               &  $W_r$ &  &                & $\Delta v$ & $\rho $   & $s$ \\
$z_{\rm{abs}}$ &   (\AA )  & Des & $z_{\rm{gal}}$ & (\kms )    & (\h ) & (\mpc
) & $M_{\rm{lim}}$ \\
\hline
0.0245 & 0.31 & Obj192   & 0.0238      & $-205$   & 578.4  & 2.06 & $-15.8$ \\ 
0.0300 & 0.05 & \multicolumn{3}{c}{\it no galaxy} & $>745$ & ... & $-16.3$  \\ 
0.0399 & 0.18 & \multicolumn{3}{c}{\it no galaxy} & $>975$ & ... & $-16.9$  \\ 
0.0411 & 0.04 & \multicolumn{3}{c}{\it no galaxy} & $>1002$& ... & $-16.9$  \\ 
0.0454 & 0.12 & \multicolumn{3}{c}{\it no galaxy} & $>1099$& ... & $-17.2$  \\ 
0.0570 & 0.06 & Obj216   & 0.0568      & $-48$    & 1318.9 & 1.42 & $-17.7$ \\ 
0.0672 & 0.12 & \multicolumn{3}{c}{\it no galaxy} & $>1569$& ... & $-18.1$  \\ 
0.1113 & 0.09 & Obj242   & 0.1116      & 67       & 563.2  & 0.89 & $-19.2$ \\ 
0.1197 & 0.04 & Obj172   & 0.1216      & 508      & 104.0  & 4.28 & $-19.3$ \\
0.1214 & 0.63 & Obj172   & 0.1216      & 54       & 104.0  & 0.46 & $-19.4$ \\
\hline
\end{tabular}
\end{minipage}
\end{table*}

\begin{figure}
\centerline{\psfig
{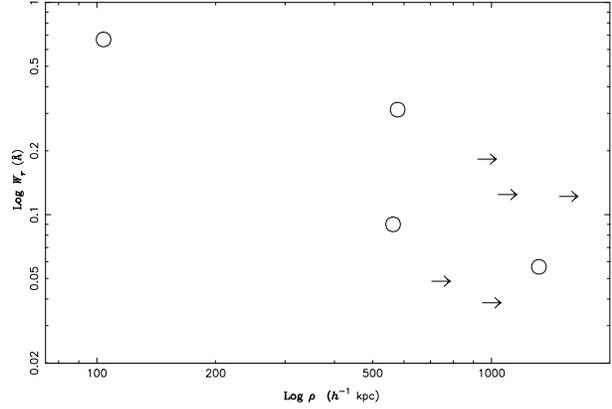}}
\caption{Plot of the distribution of rest equivalent
width, $W_r$, of the \lya\ lines towards Q1821+643, 
{\it vs.}\ impact parameters, $\rho $,  of the nearest galaxies within
$\Delta v\:=\:\pm600$~\kms\ of the \lya\ system.
Open circles represent detections of a galaxy; 
lower limits correspond to the size that 30 arcmin corresponds to at the
redshift of the absorption system.
}
\label{fig:percent2} 
\end{figure}

Of the ten \lya\ lines listed in Table~4 we are unable to detect any galaxies
at similar redshifts from five (50\%) of them.  This result is shown
graphically in a plot of $W_r$ vs.\ $\rho$ for nine of the absorption systems
in Fig.~\ref{fig:percent2} (for this figure, we have assumed that the weak
line at $z_{\rm{abs}}\:=\:0.1197$, separated by 450~\kms\ from the
$z_{\rm{abs}}\:=\:0.1214$ line, is
associated with the galaxy 104~\h\ from the QSO sightline).  The figure
suggests that at least half of the \lya\ lines are intergalactic in origin.  As
mentioned above, eight of the ten \lya\ lines are weaker than
$\simeq\:0.2$~\AA ; hence we are discussing a population of \lya\ lines much
weaker than those detectable in FOS spectra.  Six of these eight weak
absorbers have no bright galaxies at separations less than ($0.7-1.3)$~\mpc .
There are two strong lines ($W_r\:>\:0.3$~\AA) at $z_{\rm{abs}}\:=\:0.0245$
and $0.1214$, and both have galaxies 578 and 104~\h\ from the QSO sightline
respectively. Again, the question is whether these associations are
coincidental, a question we attempt to answer in \S3.3.

Our survey detects one obvious group of galaxies at $z\approx\:0.05$ (see
Fig.~\ref{fig:wedges}).  As noted above, possible absorption by \lya\ at
this redshift is unfortunately obscured by a weak Galactic C~I absorption
line. It is clear though that strong \lya\ absorption, with
equivalent widths greater than the $W$(C~I$\lambda
1277$)$\:\simeq\:0.1$~\AA\ observed, is {\it not} associated with this group
of galaxies. A direct association between galaxy groups and \lya\ absorption
systems has been claimed by Lanzetta~\etal\ (1996) for a group of galaxies
toward Q1545+2101 at $z\:=\:0.26$; in their example, all five galaxies
constituting the group are within $\rho\:=\:500$~\h\ of the QSO sightline,
and the resulting \lya\ absorption is strong ($W\:=\:0.8$~\AA )
and broad. In our data, there are 12 galaxies in a redshift interval
$0.04950-0.05250$ (which corresponds to depth of 8~\mpc\ along the line of
sight) but all are at separations of $\rho\:=\:0.44-1.01$~\mpc . This
clearly shows that loose groups of galaxies --- where intracluster gas might
be expected to produce absorption --- need not produce \lya\
absorption. M93 found a similar example towards 3C~273.

In contrast, there are three absorption lines at
$z_{\rm{abs}}\:=\:0.0399-0.0454$ (a depth of $\sim\:15$~\mpc ) which are
offset from this galaxy group, and unassociated with any galaxies brighter
than $M\:\simeq\:-17.0$.  There are $2-3$ published examples of a \lya\ line
lying in a void between galaxy structures at low-redshift (Shull~\etal\ 1996),
and, if the identification of the $z_{\rm{abs}}\:=0.0399$ and 0.0454 systems
towards Q1821+643 is correct (see \S2.2), then this would be another good
example of \lya\ lines avoiding galaxies.  Of course, at $z\:\simeq\:0.04$, 30
arcmins corresponds to only 1\mpc , which makes it possible that galaxies
forming part of a loose group at these redshifts might be missed.  If we
consider galaxies listed in the NED outside our 30 arcmin survey radius, we
find only eight nearby galaxies, at large distances of $3.0-10.0$~\mpc\ from
the QSO line of sight within a redshift range $z\:=\:0.038-0.042$. Although
the catalogue of galaxies from the NED is in no sense complete, (and drawing
any inferences from such a catalog may be highly misleading), this does
suggest that some \lya\ absorbers avoid galaxies on scales of at least
a few Mpc.

\subsection{Galaxy-absorber correlation functions: are absorbers associated
with galaxies?}

It is clear from the proceeding section that we need a statistical approach to
establishing whether apparent galaxy-absorber correlations are real.  In this
section, therefore, we calculate the galaxy-galaxy and galaxy-absorber
cross-correlation functions in our data set to test the significance of such
associations.

We have used the formalism given by Mann, Saunders \& Taylor (1996), which for
the galaxy-galaxy correlation function reduces to the estimator of
$\xi_{\rm{gg}}$ given by Hamilton (1993).  To calculate the galaxy-absorber
cross-correlation, $\xi_{\rm{gl}}$, in redshift space, where separations are
denoted as $s$, we have used their generalization

\begin{equation}
1+\xi_{\rm{gl}} (s) = \frac{D_g D_l(s)}{D_g R_l (s)} \frac{R_g R_l 
(s)}{R_g D_l (s)}
\end{equation}

where $D_g D_l(s)$ is the weighted pair counts of galaxies and lines whose
separation places them in a bin centred on $s$, $R_g R_l$ is the corresponding
weighted count for a random catalogue of a) galaxies covering the same area of
sky and having the same selection function as the data, and b) lines randomly
distributed along the line of sight with a correction for an evolving number
density with redshift. $D_g R_l$, $R_g D_l$ and $R_g R_l$ are the weighted
counts of cross pairs between the data and the random catalogs.

For the weighting of the pair counts we adopt the now `standard' choice of
using either $w(r)\:=\:1$ --- no weighting ---
(where $r$ denotes the radial distance of a
galaxy/line from the observer) or, for galaxies,

\begin{equation}
w(r,s)\:=\:1/[1+4\pi \phi(r) J_3(s)]
\end{equation}

(Peebles 1973; Hamilton 1993; Mann, Saunders \& Taylor 1996; Tucker~\etal\
1997).  Here, $\phi(r)$ is the selection function, which we take simply to be 
the
number density of galaxies in our survey, $n(r)/V(r)$, as calculated from our
data. In this weighting scheme, $J_3(s)$ is defined as

\begin{equation}
J_3(s)\:\equiv\:\int^s_0 x^2 \xi(x) dx
\end{equation}

and we have used the results from the Las Campanas Redshift Survey (LCRS;
Tucker~\etal\ 1996), $\xi(s) = (s_0/s)^\gamma$ with $s_0\:=\:6.3$ and $\gamma
= 1.52$ to evaluate $J_3(s)$. In fact, the correlation function we 
calculate is relatively insensitive to using this weighting function.
 For absorption lines, $w(r)$ is always set
equal to unity. To ensure the correct number density of lines, random
catalogs were generated with the number of lines
$n(z)\:\propto\:(1+z)^\gamma$, where $\gamma\:=\:0.42$ (Bahcall~\etal\ 
1996). Although this is inevitably only
 an approximation, because we have included \lya\ lines with
equivalent widths less than the $0.24$~\AA\ used to evaluate $n(z)$, the
correction has little effect over the small redshift range discussed herein.
For all estimates of $\xi_{gg}$ and $\xi_{gl}$, we have used 10000 points for
the random catalogues, and a maximum redshift of $z\:=\:0.2$.
For $\xi_{gl}$, we have used a minimum redshift of 0.0060 for
lines and galaxies, since below this, we cannot distinguish \lya\ lines from
the Damped \lya\ line of our own Milky Way.

\begin{figure}
\centerline{\psfig
{figure=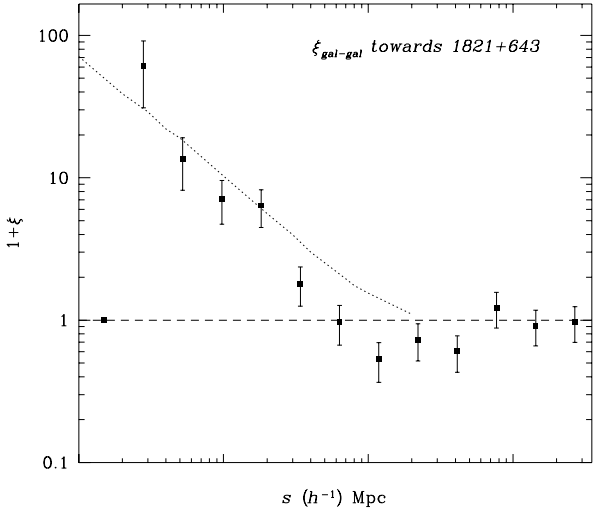,height=8cm,angle=270}}
\caption{The galaxy auto-correlation function, $\xi_{\rm{gg}}$, 
of galaxies in our sample
towards Q1821+643. For comparison, the
auto-correlation of galaxies found from 
the Las Camapanas Redshift Survey (Tucker~\etal\ 1997) 
is shown as a dotted line.}
\label{fig:cross1} 
\end{figure}

We first show in Fig.~\ref{fig:cross1} (top) 
the galaxy auto-correlation function
for galaxies weighted in the way described in equation 2. Errors shown are
$1\sigma$ error bars calculated in the way prescribed by Mo~\etal\ (1992).
Also plotted is $1+\xi_{\rm{gg}}$ from the LCRS (dotted
line). The figure shows, unsurprisingly, that the galaxies are clustered in
much the same way as seen in more extensive galaxy surveys. 

\begin{figure}
\vspace*{-5mm}
{\psfig
{figure=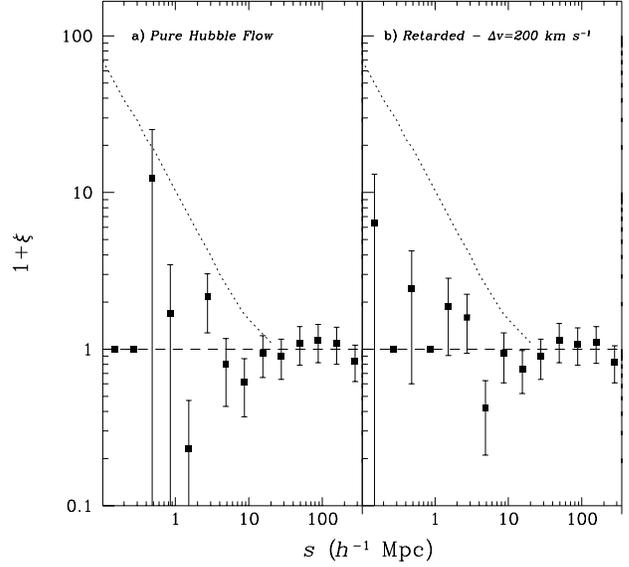,height=10cm,angle=270}}
\vspace*{-1cm}
\caption{The galaxy-absorber cross-correlation function, $\xi_{\rm{gl}}$, 
towards
Q1821+643. $\xi_{\rm{gl}}$ is calculated (a) assuming that values of $\Delta v$
between absorbers and galaxies arise from the Hubble Flow, and 
(b) assuming a retarded Hubble Flow such that $\Delta v$ is set to zero if
$\Delta v\:\leq\:200$~\kms .
For comparison, the galaxy-galaxy correlation
function found by Tucker~\etal\ (1997) for the LCRS is again 
shown as a dotted line.}
\label{fig:cross2} 
\end{figure}

In Fig.~\ref{fig:cross2} we plot the galaxy-absorber correlation function,
$\xi_{\rm{gl}}$.  For bins with no galaxies or \lya\ clouds detected, we have
set $\xi_{\rm{gl}}\:=\:1$ to indicate the size of bins used to calculate
$\xi_{\rm{gl}}$. In Fig.~\ref{fig:cross2}a the correlation is dominated by a
single point between $0.36-0.64$~\mpc ; this is Obj172 lying 104~\h\ from the
QSO sightline (Table~\ref{neargals}) at a redshift of 0.1216.  This
correlation calculation assumes a (comoving) Hubble flow; hence the difference
between $z_{\rm{gal}}$ and $z_{\rm{abs}}$ of $\simeq\:0.0002$ means that the
distance of galaxy and absorber becomes 518~\h , even though this value of
$\Delta z$ is within our measurement errors of $z_{\rm{gal}}$ and
$z_{\rm{abs}}$.  Nevertheless, the detection of a galaxy at these impact
parameters remains statistically significant; given a random distribution of
galaxies and absorbers, we would only have expected 0.02 pairs instead of the
one we found.  Hence this galaxy-absorber association is unlikely to have
arisen by chance.  Apart from this pair, Fig.~\ref{fig:cross2}a shows that the
majority of absorbers are uncorrelated with galaxies, that is, that
galaxies at similar redshifts to absorbers on scales of $0.5-2$~\mpc\ in
Fig.~\ref{fig:percent} probably arise by chance. We note a slight excess
between $s\:=\:2.0-3.6$~\mpc , but the point is within $2\sigma$ of
$\xi_{\rm{gl}}\:=\:0$ (7 pairs were expected, 10 were detected).  We note that
the lack of any galaxy-absorber pairs at separations of $s\:=\:0.15$ and
0.27~\mpc\ is not significant: compared with the randomly generated catalogue
of galaxies and absorbers, less than 1 galaxy-absorber association was
expected for each of these bins.

The problems in taking the projected two-point correlation function and
estimating the proper three-dimensional spatial correlation function (Davis \&
Peebles 1983) for galaxies and absorbers are discussed in detail by M93.  It
is of more interest, however, to calculate $\xi_{\rm{gl}}$ using a ``retarded
Hubble Flow'' to take into account the possibility that a galaxy and an
absorber may be at the same (radial) distance from us, but have peculiar
velocities with respect to each due to the origin of the absorbing gas
(e.g. halo gas co-rotating with a galaxy disk). Hence in
Fig.~\ref{fig:cross2}b we calculate $\xi_{\rm{gl}}$ with $\Delta v$ set to zero
if the measured value is $|\Delta v|\:\leq\:200$~\kms , and $\Delta v$
corrected to $\Delta v - 200$~\kms\ if $|\Delta v|\:>\:200$~\kms . (The
resulting correlation is changed little by adopting values of $100-300$~\kms\
as the criterion for setting $\Delta v\:=\:0$.)  The figure shows that in 
this case, absorbers
and galaxies tend to more consistently have values of
$\xi_{\rm{gl}}\:>\:0$. This suggests that absorbers and galaxies are weakly
associated if it is {\it assumed} that the velocity difference is not related
to any real physical separation, i.e. if the real separation between galaxy
and absorber is the measured value of $\rho$. However, it is important to
realise not only that all the points are within $2\sigma$ of
$\xi_{\rm{gl}}\:=\:0$, but that this correlation is sensitive to the exact
sample of \lya\ lines used; simply removing a line at random from the list to
test the robustness of the correlation can remove the apparent consistency of
having most of the points with 
$\xi_{\rm{gl}}\:>\:0$. This lack of robustness is not surprising given
the small sample size of our data set.

We conclude that weak absorbers are uncorrelated, or at best, weakly
correlated, with galaxies, on scales of beyond a few hundred \h\ or more.
Given that there may be more of a significance for the retarded Hubble Flow
estimate of $\xi_{\rm{gl}}$, we can speculate on what this might mean. In what
scenarios could galaxies and absorbers be at the same radial distance from us?
One possibility is that absorbing gas is directly associated with the
identified galaxy. This seems unlikely for points in
Fig.~\ref{fig:percent} and~\ref{fig:percent2} with $\rho\:=\:0.5-2$~\mpc\
since there is good evidence that individual galaxy halos do not have
absorbing gas beyond $\rho\:=\:200-300$~\h\ (LBTW; BBP). A second possibility
is that the gas is a remnant of prior galaxy-galaxy interactions. Although
tidal debris is not a favoured explanation for the origin of the
$z\:\sim\:0.5$ Mg~II absorption lines (see, e.g. Steidel~\etal\ 1994), nearby
interacting galaxies have been shown to produce the characteristics of higher
redshift Mg~II systems (Bowen~\etal\ 1995), while tidal debris as an origin of
\lya\ absorbers has been discussed by Morris \& van den Bergh
(1994). This interpretation will remain hard to test, as observing the more
subtle signs of galaxy-galaxy interactions is difficult.

Finally, galaxies and absorbers could be at similar distances if they inhabit
the same dark matter (DM) structures which constitute the large scale
structure (LSS) of the Universe. Many CDM-based numerical simulations
(e.g. Cen~\etal\ 1994; Petitjean, M\"{u}cket \& Kates 1995; Zhang~\etal\ 1995;
M\"{u}cket~\etal\ 1996) now point to the possibility that low column density
H~I gas simply follows the same LSS as galaxies (at least at high redshift).
It seems likely that galaxies and absorbers which lie in DM sheets and
filaments will  have small (but non-zero) velocity differences, since the
velocity width of galaxy shells seen in local LSS is $200-300$~\kms\
(e.g. Santiago 1995). Hence the measured values of $\rho$ may well represent
real distances between galaxies and absorbers, at least for sheets aligned
perpendicular to our line of sight.

\subsection{Conclusions and Summary}
   
We have detected a single galaxy 104~\h\ from the line of sight of 
Q1821+643; \lya\ absorption arises at the same redshift as 
the galaxy to within measurement errors.  The association is statistically 
significant compared to a randomly distributed ensemble of clouds and 
galaxies, suggesting a direct link between absorber and galaxy.  The line is 
strong ($W_r\:=\:0.63$~\AA ) and its detection is consistent with 
those of similarly strong lines within 160~\h\ of bright galaxies found 
by LBTW.  The next nearest galaxies are 291~\h\ and 410~\h\ away from the 
sightline, and these show no absorption to very low equivalent width limits 
($W_r\:<0.05$~\AA ) within 600~\kms .  Beyond 500~\h , several absorption 
systems are found at similar redshifts to the galaxies, but these 
coincidences are likely to be accidental.

Five of ten (50\%) \lya\ absorption systems out to a redshift of $0.12$
have no galaxies within 600~\kms\ and separations less than ($0.7-1.3$)~\mpc .
At these redshifts, our magnitude limited redshift survey would have been able
to detect at least $L^*$ luminosity galaxies. Excluding the one \lya\ system
apparently 
associated with the galaxy described above, two \lya\ systems have
galaxies within 600~\kms\ and $\rho \:=\:0.4-0.7$~\mpc, but these associations
are again probably coincidental.  Eight of the ten lines have {\it weak}
equivalent widths ($W_r\:<\:0.2$~\AA), and we conclude that this population is
uncorrelated, or at best, weakly correlated, with galaxies. A similar result
was found by Stocke~\etal\ (1995) for lines with $W_r\:\leq\:0.1$~\AA\
detected in their own and previously published data sets. 
 
The nature of the strong lines is less clear.  The detection of strong 
lines within 200~\h\ of galaxies by LBTW, and the association of one of two 
strong lines (out to a redshift of 0.12) with a galaxy at a separation of 
104~\h\ in our own survey supports the idea that strong lines are 
associated with galaxies while weak ones are intergalactic in origin.  This 
is the `two-population' explanation proposed by Mo \& Morris (1994) and 
Bahcall~\etal\ (1996).  The difficulty in unambiguously drawing this 
conclusion from our analysis is that our data set contains too few strong 
lines at low redshift to adequately characterise the nature of this 
population.  Le Brun~\etal\ (1996) 
have published the only other available survey 
of the association of strong (FOS-detectable) lines with galaxies;  of 
13 \lya\ systems listed with $W_r\:>\:0.24$~\AA, four have galaxies 
coincident in velocity which are within 200~\h\ of a QSO sightline,
five are in the range 
$\rho  \:=\:200-500$~\h , and the remaining four are at separations beyond 
1.0~\mpc\ --- a roughly even spread in impact parameters.  Much of this 
ambiguity may be down to the 43~\% completeness of their survey at 
$m_r\:=\:18.5$, which ensures considerable difficulty in detecting 
intermediate brightness galaxies at the redshifts of the \lya\ systems 
studied, $z_{\rm{abs}}\:=\:0.09-0.77$. However, detection of even a 
few bright galaxies so close to \lya\ systems may suggest that strong 
lines are directly associated with galaxies. Clearly, to explore this 
possibility further, more complete galaxy surveys are required at 
the redshifts of strong lines. A second approach would be 
to directly search for \lya\ absorption in halos of nearby, isolated 
galaxies at impact parameters of $0-500$~\h .

Our results do not rule out the possibility that strong \lya\ lines are 
directly associated with galaxies of {\it all } luminosities --- including 
faint dwarf galaxies which have not been found in the surveys described 
above --- on scales of a few hundred \h. This would explain why it has 
been easy to find some absorbing galaxies (i.e. the bright ones) while 
some strong lines appear to have no associated galaxies (those which 
are too faint to be detected).  
Only Rauch \etal\ (1996) have searched for low surface brightness galaxies 
which might be responsible for \lya\ absorption lines toward 3C~273, and 
none were detected, although two faint galaxies (CGCG014$-$054, 
$M_B\:=\:-16.2$ and the H~I cloud H~I~1225+01, $M_B\:=\:-14.0$) are at 
similar redshifts to the lowest redshifted \lya\ lines at 1012 and 1582~\kms\ 
and within $\rho \:=\:200$~\h .  BBP also found \lya\ absorption toward 
Q1001+2910 at the same velocity as UGCA201 ($M_B\:=\:-16.2$, $\rho  
\:=\:117$~\h ) with no other galaxies nearby.  Unfortunately, it is hard to 
test whether these coincidences are accidental or not.

\bigskip

The amount of data needed to fully understand the origin
of the \lya -forest is still far from available. Obtaining high quality, high
resolution \hst\ spectra toward Q1821+643 demonstrates the importance of such
data, not only for exploring the weak population of \lya\ lines, but simply in
disentangling the \lya -forest from Galactic absorption lines when working at
low-redshift. The nature of the strong lines remains unclear, and as mentioned
above, a direct search for \lya\ lines within a few hundred \h\ of nearby,
isolated galaxies would provide much needed information as to whether strong
lines alone are directly associated with galaxies.

\section*{ACKNOWLEDGMENTS}

It is a pleasure to thank Andy Taylor, Licia Verde, and John
Peacock for highly illuminating discussions.  We are also very grateful to
Mike Irwin for providing candidate objects in the field of Q1821+643 from APM
scans. IRAF is distributed by the National Optical Astronomy Observatories,
which is operated by the Association of Universities for Research in
Astronomy, Inc., under cooperative agreement with the National Science
Foundation. The NASA/IPAC Extragalactic Database (NED) is operated by the Jet
Propulsion Laboratory, Caltech, under contract with National Aeronautics and
Space Administration.

\end{document}